\providecommand{\paper}[1]{#1}

\paper{
\newcommand{\thesis}[1]{}

\documentclass[a4paper]{article}
\usepackage{amsmath}
\usepackage{amssymb}
\usepackage{xy}
\usepackage[thmmarks]{ntheorem}

\frenchspacing \setlength{\parindent}{0pt} \setlength{\parskip}{1ex
plus 0.5ex minus 0.2ex}

\author{Robert Brijder \and Hendrik Jan Hoogeboom \and Grzegorz Rozenberg\\
\mbox{ }\\
Leiden Institute of Advanced Computer Science, Universiteit Leiden,\\
Niels Bohrweg 1, 2333 CA Leiden, The Netherlands,\\
\texttt{rbrijder@liacs.nl}}
\title{How Overlap Determines the Macronuclear Genes in Ciliates}

\xyoption{curve} \xyoption{arrow} \xyoption{matrix}

\theoremstyle{break} \theorembodyfont{\upshape}

\theoremsymbol{}

\newtheorem{Theorem}{Theorem}
\newtheorem{Lemma}[Theorem]{Lemma}
\newtheorem{Corollary}[Theorem]{Corollary}
\newtheorem{Example}{Example}

\theoremsymbol{\rule{1ex}{1ex}}
\newtheorem{Definition}[Theorem]{Definition}

\theoremstyle{nonumberbreak}
\newtheorem{Proof}{Proof}

\theoremsymbol{}
\newtheorem{Remark}{Remark}

\begin{document}

\date{}
\maketitle

\newcommand{\pset}[1]{{\mathbf #1}}
\newcommand{\fredgr}{\mathcal{R}}
\newcommand{\RGVertL}[1]{I_{#1}}
\newcommand{\RGVertR}[1]{I'_{#1}}
\newcommand{\redge}{\ar@2{-}}
\newcommand{\dedge}{\ar@{-}}

\newcommand{\overlapgr}{\gamma}
\newcommand{\overlapgru}[1]{\gamma_{#1}}
\newcommand{\compress}{\mathrm{cps}}
\newcommand{\rspos}{{r \! spos}}
\newcommand{\posn}{\mathrm{posn}}
\newcommand{\dom}{\mathrm{dom}}
\newcommand{\posve}{\mathrm{pos}}
\newcommand{\negve}{\mathrm{neg}}
\newcommand{\RGOVertNonRoot}[1]{J_{#1}}
\newcommand{\RGOVertRoot}[1]{J'_{#1}}
\renewcommand{\emptyset}{\varnothing}

}

\thesis{
\chapter{How Overlap Determines the Macronuclear Genes in Ciliates}
}

\begin{abstract}
Formal models for gene assembly in ciliates have been developed, in
particular the string pointer reduction system (SPRS) and the graph
pointer reduction system (GPRS). The reduction graph is a valuable
tool within the SPRS, revealing much information about how gene
assembly is performed for a given gene. The GPRS is more abstract
than the SPRS and not all information present in the SPRS is
retained in the GPRS. As a consequence the reduction graph cannot be
defined for the GPRS in general, but we show that it can be defined
(in an equivalent manner as defined for the SPRS) if we restrict
ourselves to so-called realistic overlap graphs. Fortunately, only
these graphs correspond to genes occurring in nature. Defining the
reduction graph within the GPRS allows one to carry over several
results within the SPRS that rely on the reduction graph.
\end{abstract}


\section{Introduction}
Gene assembly is a biological process that takes place in a large
group of one-cellular organisms called ciliates. The process
transforms one nucleus, called the micronucleus, through a large
number of splicing operations into another nucleus, called the
macronucleus. The macronucleus is very different from the
micronucleus, both functionally and in terms of differences in DNA.
Each gene occurring in the micronucleus in transformed into a
corresponding gene in the macronucleus. Two models that are used to
formalize this process are the string pointer reduction system
(SPRS) and the graph pointer reduction system (GPRS). The former
consist of three types of string rewriting rules operating on
strings, called legal strings, while the latter consist of three
types of graph rewriting rules operating on graphs, called overlap
graphs. The GPRS can be seen as an abstraction of the SPRS, however
it is not fully equivalent with the SPRS: some information present
in the SPRS is lost in the GPRS.

Legal strings represent genes in their micronuclear form. The
reduction graph, which is defined for legal strings, is a notion
that describes the corresponding gene in its macronuclear form
(along with its waste products). Moreover, it has been shown that
the reduction graph retains much information on which string
negative rules (one of the three types of string rewriting rules)
can be or are used in this transformation
\cite{Extended_paper,DBLP:conf/complife/BrijderHR05,DBLP:conf/complife/BrijderHR06}.
Therefore it is natural to define an equivalent notion for the GPRS.
However, as we will show, since the GPRS loses some information
concerning the application of string negative rules, there is no
unique reduction graph for a given overlap graph. We will show
however, that when we restrict ourselves to `realistic' overlap
graph then there is a unique reduction graph corresponding to this
graph. These overlap graphs are called realistic since non-realistic
overlap graphs cannot correspond to (micronuclear) genes. Moreover,
we explicitly define the notion of reduction graph for these overlap
graphs (within the GPRS) and show the equivalence with the
definition for legal strings (within the SPRS). Finally, we show
some immediate results due to this equivalence, including an open
problem formulated in Chapter~13 in \cite{GeneAssemblyBook}.

In Section~\ref{sect_notation} we recall some basic notions and
notation concerning sets, strings and graphs. In
Section~\ref{sect_gene_assembly} we recall notions used in models
for gene assembly, such as legal strings, realistic strings and
overlap graphs. In Section~\ref{sect_reduction_graph} we recall the
notion of reduction graph within the framework of SPRS and we prove
a few elementary properties of this graph for legal strings. In
particular we establish a calculus for the sets of overlapping
pointers between vertices of the reduction graph. In
Section~\ref{sect_reduction_graph_realistic} we prove properties of
the reduction graph for a more restricted type of legal strings, the
realistic strings. It is shown that reduction graphs of realistic
strings have a subgraph of a specific structure, the root subgraph.
Moreover the existence of the other edges in the reduction graph is
shown to depend directly on the overlap graph, using the calculus
derived in the Section~\ref{sect_reduction_graph}. In
Section~\ref{sect_compress_function} we provide a convenient
function for reduction graphs (but not only reduction graphs) which
simplifies reduction graphs without losing any information. In
Section~\ref{sect_overlap_to_red_graph} we define the reduction
graph for realistic overlap graphs, and prove the main theorem of
this paper: the equivalence of reduction graphs defined for
realistic strings and reduction graphs defined for realistic overlap
graphs. In Section~\ref{sect_consequences} we show immediate
consequences of this theorem.

\section{Notation and Terminology} \label{sect_notation}
In this section we recall some basic notions concerning functions,
strings, and graphs. We do this mainly to set up the basic notation
and terminology for this paper.

The cardinality of set $X$ is denoted by $|X|$. The symmetric
difference of sets $X$ and $Y$, $(X \backslash Y) \cup (Y \backslash
X)$, is denoted by $X \oplus Y$. Being an associative operator, we
can define the symmetric difference of a family of sets $(X_i)_{i
\in A}$ and denote it by $\bigoplus_{i \in A} X_i$. The
\emph{composition} of functions $f: X \rightarrow Y$ and $g: Y
\rightarrow Z$ is the function $g f: X \rightarrow Z$ such that $(g
f) (x) = g(f(x))$ for every $x \in X$. The restriction of $f$ to a
subset $A$ of $X$ is denoted by $f|A$.

We will use $\lambda$ to denote the empty string.
For strings $u$ and $v$, we say that $v$ is a \emph{substring of
$u$} if $u = w_1 v w_2$, for some strings $w_1$, $w_2$; we also say
that $v$ \emph{occurs in $u$}. Also, $v$ is a \emph{cyclic substring
of $u$} if either $v$ is a substring of $u$ or $u = v_2 w v_1$ and
$v = v_1 v_2$ for some strings $v_1, v_2, w$.
We say that $v$ is a \emph{conjugate of $u$} if $u = w_1 w_2$ and $v
= w_2 w_1$ for some strings $w_1$ and $w_2$. For a string $u = x_1
x_2 \cdots x_n$ over $\Sigma$ with $x_i \in \Sigma$ for all $i \in
\{1,\ldots,n\}$, we say that $v = x_n x_{n-1} \cdots x_1$ is the
\emph{reversal of $u$}. A \emph{homomorphism} is a function
$\varphi: \Sigma^* \rightarrow \Delta^*$ such that $\varphi(uv) =
\varphi(u) \varphi(v)$ for all $u,v \in \Sigma^*$.

We move now to graphs. A \emph{labelled graph} is a 4-tuple
$$
G = (V,E,f,\Gamma),
$$
where $V$ is a finite set, $E \subseteq \{ \{x,y\} \mid x,y \in V, x
\not= y \}$, and $f: V \rightarrow \Gamma$.

The elements of $V$ are called \emph{vertices} and the elements of
$E$ are called \emph{edges}. Function $f$ is the \emph{labelling
function} and the elements of $\Gamma$ are the \emph{labels}. We say
that $G$ is \emph{discrete} if $E = \emptyset$. Labelled graph $G' =
(V',E',f|V',\Gamma)$ is a \emph{subgraph of $G$} if $V' \subseteq V$
and $E' \subseteq E_{V'} = E \cap \{ \{x,y\} \mid x,y \in V', x
\not= y \}$. If $E' = E_{V'}$, we say that $G'$ is the
\emph{subgraph of $G$ induced by $V'$}.

A string $\pi = e_1 e_2 \cdots e_n \in E^*$ with $n \geq 1$ is a
\emph{path in $G$} if there is a $v_1 v_2 \cdots v_{n+1} \in V^*$
such that $e_i = \{v_i, v_{i+1}\}$ for all $1 \leq i \leq n$.
Labelled graph $G$ is \emph{connected} if there is a path between
every two vertices of $G$. A subgraph $H$ of $G$ induced by $V_H$ is
a \emph{component of $G$} if both $H$ is connected and for every
edge $e \in E$ we have either $e \subseteq V_H$ or $e \subseteq V
\backslash V_H$.

As usual, labelled graphs $G = (V,E,f,\Gamma)$ and $G' =
(V',E',f',\Gamma)$ are \emph{isomorphic}, denoted by $G \approx G'$,
if there is a bijection $\alpha: V \rightarrow V'$ such that $f(v) =
f'(\alpha(v))$ for $v \in V$, and
$$
\{x,y\} \in E \mbox{ iff } \{\alpha(x),\alpha(y)\} \in E'
$$
for $x,y \in V$. Bijection $\alpha$ is then called an
\emph{isomorphism from $G$ to $G'$}.

In this paper we will consider graphs with two sets of edges.
Therefore, we need the notion of 2-edge coloured graphs. A
\emph{2-edge coloured graph} is a 5-tuple
$$
G = (V,E_1,E_2,f,\Gamma),
$$
where both $(V,E_1,f,\Gamma)$ and $(V,E_2,f,\Gamma)$ are labelled
graphs.

The basic notions and notation for labelled graphs carry over to
2-edge coloured graphs. However, for the notion of isomorphism care
must be taken that the two sorts of edges are preserved. Thus, if $G
= (V,E_1,E_2,f,\Gamma)$ and $G' = (V',E'_1,E'_2,f',\Gamma')$ are
2-edge coloured graphs, then it must hold that for an isomorphism
$\alpha$ from $G$ to $G'$,
$$
(x,y) \in E_i \mbox{ iff } (\alpha(x),\alpha(y)) \in E_i'
$$
for $x,y \in V$ and $i \in \{1,2\}$.

\section{Gene Assembly in Ciliates} \label{sect_gene_assembly}
Two models that are used to formalize the process of gene assembly
in ciliates are the string pointer reduction system (SPRS) and the
graph pointer reduction system (GPRS). The SPRS consist of three
types of string rewriting rules operating on \emph{legal strings}
while the GPRS consist of three types of graph rewriting rules
operating on \emph{overlap graphs}. For the purpose of this paper it
is not necessary to recall the string and graph rewriting rules; a
complete description of SPRS and GPRS, as well as a proof of their
``weak'' equivalence, can be found in \cite{GeneAssemblyBook}. We do
recall the notions of legal string and overlap graph, and we also
recall the notion of realistic string.

We fix $\kappa \geq 2$, and define the alphabet $\Delta =
\{2,3,\ldots,\kappa\}$. For $D \subseteq \Delta$, we define $\bar D
= \{ \bar a \mid a \in D \}$ and $\Pi_D = D \cup \bar D$; also $\Pi
= \Pi_{\Delta}$. The elements of $\Pi$ will be called
\emph{pointers}. We use the ``bar operator'' to move from $\Delta$
to $\bar \Delta$ and back from $\bar \Delta$ to $\Delta$. Hence, for
$p \in \Pi$, $\bar {\bar {p}} = p$. For $p \in \Pi$, we define
$\pset{p} =
\begin{cases} p & \mbox{if } p \in \Delta \\ \bar{p} & \mbox{if }
p \in \bar{\Delta}
\end{cases}$
, i.e., $\pset{p}$ is the ``unbarred'' variant of $p$.

For a string $u = x_1 x_2 \cdots x_n$ with $x_i \in \Pi$ ($1 \leq i
\leq n$), the \emph{complement of $u$} is $\bar x_1 \bar x_2 \cdots
\bar x_n$. The \emph{inverse of $u$}, denoted by $\bar u$, is the
complement of the reversal of $u$, thus $\bar u = \bar x_n \bar
x_{n-1} \cdots \bar x_1$. The \emph{domain of $u$}, denoted by
$dom(u)$, is $\{ \pset{p} \mid \mbox{$p$ occurs in $v$} \}$. We say
that $u$ is a \emph{legal string} if for each $p \in dom(u)$, $u$
contains exactly two occurrences from $\{p,\bar p\}$.

We define the alphabet $\Theta_{\kappa} = \{M_i, \bar M_i \mid 1
\leq i \leq \kappa \}$. We say that $\delta \in \Theta^*_{\kappa}$
is a \emph{micronuclear arrangement} if for each $i$ with $1 \leq i
\leq \kappa$, $\delta$ contains exactly one occurrence from
$\{M_i,\bar M_i\}$. With each string over $\Theta_{\kappa}$, we
associate a unique string over $\Pi$ through the homomorphism
$\pi_{\kappa}: \Theta^*_{\kappa} \rightarrow \Pi^*$ defined by:
$$
\pi_{\kappa}(M_1) = 2, \quad \pi_{\kappa}(M_{\kappa}) = \kappa,
\quad \pi_{\kappa}(M_i) = i(i+1) \quad \mbox{for } 1 < i < \kappa,
$$
and $\pi_{\kappa}(\bar M_j) = \overline{\pi_{\kappa}(M_j)}$ for $1
\leq j \leq \kappa$. We say that string $u$ is a \emph{realistic
string} if there is a micronuclear arrangement $\delta$ such that $u
= \pi_{\kappa}(\delta)$. We then say that $\delta$ is a
\emph{micronuclear arrangement for $u$}.

Note that every realistic string is a legal string. However, not
every legal string is a realistic string. For example, a realistic
string cannot have ``gaps'' (missing pointers): thus $2244$ is not
realistic while it is legal. It is also easy to produce examples of
legal strings which do not have gaps but still are not realistic ---
$3322$ is such an example. Realistic strings are most useful for the
gene assembly models, since only these legal strings can correspond
to genes in ciliates.

For a pointer $p$ and a legal string $u$, if both $p$ and $\bar p$
occur in $u$ then we say that both $p$ and $\bar p$ are
\emph{positive in $u$}; if on the other hand only $p$ or only $\bar
p$ occurs in $u$, then both $p$ and $\bar p$ are \emph{negative in
$u$}. So, every pointer occurring in a legal string is either
positive or negative in it. Therefore, we can define a partition of
$\dom(u) = \posve(u) \cup \negve(u)$, where $\posve(u) = \{ p \in
\dom(u) \mid \mbox{$p$ is positive in $u$} \}$ and $\negve(u) = \{ p
\in \dom(u) \mid \mbox{$p$ is negative in $u$} \}$.

Let $u = x_1 x_2 \cdots x_n$ be a legal string with $x_i \in \Pi$
for $1 \leq i \leq n$. For a pointer $p \in \Pi$ such that
$\{x_i,x_j\} \subseteq \{p,\bar p\}$ and $1 \leq i < j \leq n$, the
\emph{p-interval of $u$} is the substring $x_i x_{i+1} \cdots x_j$.
Substrings $x_{i_1} \cdots x_{j_1}$ and $x_{i_2} \cdots x_{j_2}$
\emph{overlap in $u$} if $i_1 < i_2 < j_1 < j_2$ or $i_2 < i_1 < j_2
< j_1$. Two distinct pointers $p,q \in \Pi$ \emph{overlap in $u$} if
the $p$-interval of $u$ overlaps with the $q$-interval of $u$. Thus,
two distinct pointers $p,q \in \Pi$ overlap in $u$ iff there is
exactly one occurrence from $\{p, \bar p\}$ in the $q$-interval, or
equivalently, there is exactly one occurrence from $\{q, \bar q\}$
in the $p$-interval of $u$. Also, for $p \in \dom(u)$, we denote
$$
O_u(p) = \{ q \in \dom(u) \mid \mbox{$p$ and $q$ overlap in $u$} \},
$$
and for $0 \leq i \leq j \leq n$, we denote by $O_u(i,j)$ the set of
all $p \in \dom(u)$ such that there is exactly one occurrence from
$\{p, \bar p\}$ in $x_{i+1} x_{i+2} \cdots x_j$. Also, we define
$O_u(j,i) = O_u(i,j)$. Intuitively, $O_u(i,j)$ is the set of $p \in
\dom(u)$ for which the the substring between ``positions'' $i$ and
$j$ in $u$ contains exactly one representative from $\{ p, \bar p
\}$, where position $i$ for $0 < i < n$ means the ``space'' between
$x_i$ and $x_{i+1}$ in $u$. For $i = 0$ it is the ``space'' on the
left of $x_1$, and for $i = n$ it is the ``space'' on the right of
$x_n$. A few elementary properties of $O_u(i,j)$ follow. We have
$O_u(i,n) = O_u(0,i)$ for $i$ with $0 \leq i \leq n$. Moreover, for
$i,j,k \in \{0, \ldots, n\}$, $O_u(i,j) \oplus O_u(j,k) = O_u(i,k)$;
this is obvious when $i < j < k$, but it is valid in general. Also,
for $0 \leq i \leq j \leq n$, $O_u(i,j) = \emptyset$ iff $x_{i+1}
\cdots x_j$ is a legal string.

\begin{Definition}
Let $u$ be a legal string. The \emph{overlap graph of $u$}, denoted
by $\overlapgru{u}$, is the labelled graph
$$
(\dom(u),E,\sigma,\{+,-\}),
$$
where
$$
E = \{ \{p,q\} \mid p,q \in \dom(u), p \not= q, \mbox{and $p$ and
$q$ overlap in $u$} \},
$$
and $\sigma$ is defined by:
$$
\sigma(p) = \begin{cases} + & \mbox{if } p \in \posve(u) \\ - &
\mbox{if } p \in \negve(u)
\end{cases}
$$
for all $p \in \dom(u)$.
\end{Definition}

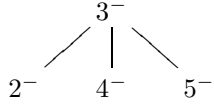
\begin{figure}
$$
\xymatrix @=15pt{
 & 3^- \ar@{-}[dl] \ar@{-}[d] \ar@{-}[dr] & \\
2^- & 4^- & 5^-
}
$$
\caption{The overlap graph of legal string $u = 24535423$.}
\label{non_realistic_overlap_graph}
\end{figure}

\begin{Example}
Let $u = 24535423$ be a legal string. The overlap graph of $u$ is
$$
\overlapgr = (\{2,3,4,5\},\{\{2,3\}, \{4,3\},
\{5,3\}\},\sigma,\{+,-\}),
$$
where $\sigma(v) = -$ for all vertices $v$ of $\overlapgr$. The
overlap graph is depicted in
Figure~\ref{non_realistic_overlap_graph}.
\end{Example}

Let $\overlapgr$ be an overlap graph. Similar to legal strings, we
define $\dom(\overlapgr)$ as the set of vertices of $\overlapgr$,
$\posve(\overlapgr) = \{ p \in \dom(\overlapgr) \mid \sigma(p) = +
\}$, $\negve(\overlapgr) = \{ p \in \dom(\overlapgr) \mid \sigma(p)
= - \}$ and for $q \in \dom(u)$, $O_{\overlapgr}(q) = \{ p \in
\dom(\overlapgr) \mid \{p,q\} \in E \} $.

An overlap graph $\overlapgr$ is \emph{realistic} if it is the
overlap graph of a realistic string. Not every overlap graph of a
legal string is realistic. For example, it can be shown that the
overlap graph $\overlapgr$ of $u = 24535423$ depicted in
Figure~\ref{non_realistic_overlap_graph} is not realistic. In fact,
one can show that it is not even \emph{realizable} --- there is no
isomorphism $\alpha$ such that $\alpha(\overlapgr)$ is realistic.

\section{The Reduction Graph} \label{sect_reduction_graph}
We now recall the (full) reduction graph, which was first introduced
in \cite{Extended_paper}.

\begin{Remark}
Below we present this graph in a slightly modified form: we omit the
special vertices $s$ and $t$, called the source vertex and target
vertex respectively, which did appear in the definition presented in
\cite{Extended_paper}. As shown in
Section~\ref{sect_reduction_graph_realistic}, in this way a
realistic overlap graph corresponds to exactly one reduction graph.
Fortunately, several results concerning reduction graphs do not rely
on the special vertices, and therefore carry over trivially to
reduction graphs as defined here.
\end{Remark}

%
%
\begin{Definition}
Let $u = p_1 p_2 \cdots p_n$ with $p_1,\ldots,p_n \in \Pi$ be a
legal string. The \emph{reduction graph of $u$}, denoted by
$\fredgr_u$, is a 2-edge coloured graph
$$
(V,E_1,E_2,f,\dom(u)),
$$
where
$$
V = \{\RGVertL{1},\RGVertL{2},\ldots,\RGVertL{n}\} \ \cup \
\{\RGVertR{1},\RGVertR{2},\ldots,\RGVertR{n}\},
$$
$$
E_{1} = \{ e_1, e_2, \ldots, e_{n} \} \mbox{ with }  e_i = \{
\RGVertR{i},\RGVertL{i+1} \} \mbox{ for } 1 \leq i \leq n-1, e_n =
\{ \RGVertR{n}, \RGVertL{1} \},
$$
\begin{eqnarray*}
E_{2} = & \{ \{\RGVertR{i},\RGVertL{j}\},
\{\RGVertL{i},\RGVertR{j}\} \mid i,j \in \{1,2,\ldots,n\}
\mbox{ with } i \not= j \mbox{ and } p_i = p_j \} \ \cup \ \\
& \{ \{\RGVertL{i},\RGVertL{j}\}, \{\RGVertR{i},\RGVertR{j}\} \mid
i,j \in \{1,2,\ldots,n\} \mbox{ and } p_i = \bar{p}_j \}, \mbox{
and}
\end{eqnarray*}
$$
\mbox{$f(\RGVertL{i}) = f(\RGVertR{i}) = \pset{p_i}$ for $1 \leq i
\leq n$.}
$$
\mbox{  }
\end{Definition}

The edges of $E_1$ are called the \emph{reality edges}, and the
edges of $E_2$ are called the \emph{desire edges}. Intuitively, the
``space'' between $p_i$ and $p_{i+1}$ corresponds to the reality
edge $e_i = \{ \RGVertR{i}, \RGVertL{i+1} \}$. Hence, we say that
$i$ is the \emph{position of $e_i$}, denoted by $\posn(e_i)$, for
all $i \in \{1,2,\ldots,n\}$. Note that positions are only defined
for reality edges. Since for every vertex $v$ there is a unique
reality edge $e$ such that $v \in e$, we also define the
\emph{position of $v$}, denoted by $\posn(v)$, as the position of
$e$. Thus, $\posn(\RGVertR{i}) = \posn(\RGVertL{i+1}) = i$ (while
$\posn(\RGVertL{1}) = n$).
%

\begin{figure}
$$
\xymatrix @=20pt{
& & \RGVertL{1},3 \ar@{-}[ddddd] & \RGVertR{1},3 \ar@{-}[dr] \ar@{-}[ddddd] & &\\
& \RGVertR{6},4 \ar@{-}@/_1.0pc/[rrrddd] \ar@{-}[ur] & & & \RGVertL{2},2 &\\
\RGVertL{6},4 \ar@{-}@/^1.0pc/[rrrrrd] & & & & & \RGVertR{2},2 \ar@{-}[d] \\
\RGVertR{5},2 \ar@{-}@/^1.0pc/[rrrrru] \ar@{-}[u] & & & & & \RGVertL{3},4\\
& \RGVertL{5},2 \ar@{-}@/_1.0pc/[rrruuu] & & & \RGVertR{3},4 \ar@{-}[dl] &\\
& & \RGVertR{4},3 \ar@{-}[ul] & \RGVertL{4},3 & &\\
}
$$
\caption{The reduction graph of $u$ of
Example~\ref{ex_red_graph_of_legal_string}.}
\label{ex_red_graph_of_legal_string_fig1}
\end{figure}
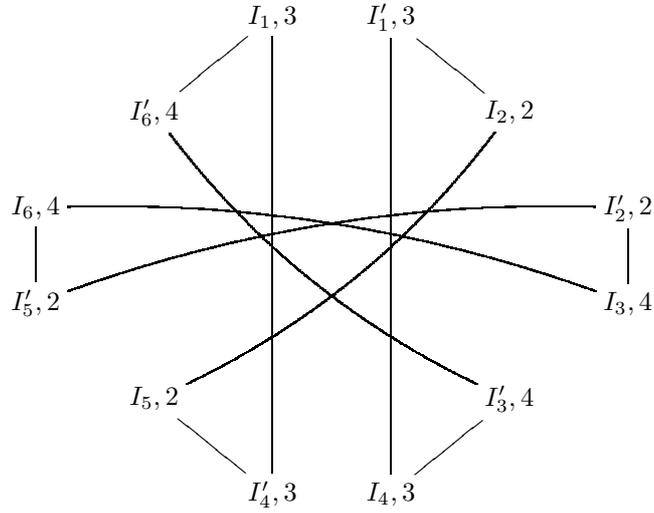

\begin{figure}
$$
\xymatrix @=20pt{
2 \dedge[d] \redge[r] & 4 \dedge[d] & 2 \dedge[d] \redge[r] & 3 \dedge[r] & 3 \redge[r] & 4 \dedge[d] \\
2 \redge[r] & 4 & 2 \redge[r] & 3 \dedge[r] & 3 \redge[r] & 4
}
$$
\caption{The reduction graph of $u$ of
Example~\ref{ex_red_graph_of_legal_string} in the simplified
representation.} \label{ex_red_graph_of_legal_string_fig2}
\end{figure}
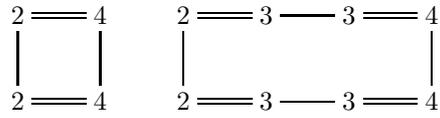

\begin{Example} \label{ex_red_graph_of_legal_string}
Let $u = 3 2 \bar 4 3 \bar 2 4$ be a legal string. Since $\bar 4 3
\bar 2$ can not be a substring of a realistic string, $u$ is not
realistic. The reduction graph $\fredgr_u$ of $u$ is depicted in
Figure~\ref{ex_red_graph_of_legal_string_fig1}. The labels of the
vertices are also shown in this figure. Note the desire edges
corresponding to positive pointers (here $2$ and $4$) cross (in the
figure), while those for negative pointers are parallel. Since the
exact identity of the vertices in a reduction graph is not essential
for the problems considered in this paper, in order to simplify the
pictorial representation of reduction graphs we will omit this in
the figures. We will also depict reality edges as ``double edges''
to distinguish them from the desire edges.
Figure~\ref{ex_red_graph_of_legal_string_fig2} shows the reduction
graph in this simplified representation.
\end{Example}

\begin{figure}
$$
\xymatrix @=30pt{
3 \dedge[r] & 3 \redge[r] & 6 \dedge[r] & 6 \redge[r] & 2 \dedge[r] & 2 \redge[d] \\
7 \redge[u] & 7 \dedge[l] & 5 \redge[l] & 5 \dedge[l] & 4 \redge[l] & 4 \dedge[l] \\
2 \dedge[r] & 2 \redge[r] & 3 \dedge[r] & 3 \redge[r] & 4 \dedge[r] & 4 \redge[d] \\
7 \redge[u] & 7 \dedge[l] & 6 \redge[l] & 6 \dedge[l] & 5 \redge[l]
& 5 \dedge[l]}
$$
\caption{The reduction graph of $u$ of
Example~\ref{ex_red_graph_of_realistic_string}.}
\label{ex_red_graph_of_realistic_string_fig1}
\end{figure}
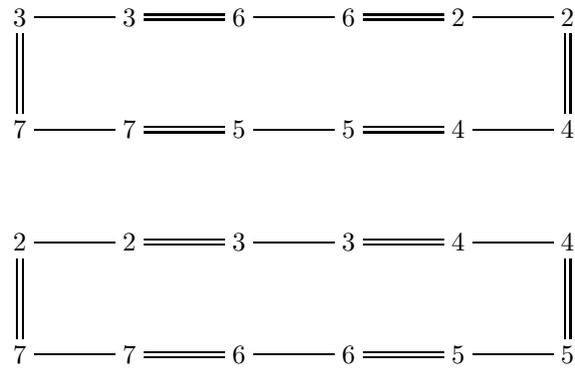

\begin{Example} \label{ex_red_graph_of_realistic_string}
Let $u = \pi_7(M_7 M_1 M_6 M_3 M_5 \overline{M_2} M_4)= 72673456
\bar 3 \bar 2 45$. Thus, unlike the previous example, $u$ is a
realistic string. The reduction graph is given in
Figure~\ref{ex_red_graph_of_realistic_string_fig1}. As usual, the
vertices are represented by their labels.
\end{Example}

The reduction graph is defined for legal strings. In this paper, we
will show how to directly construct the reduction graph of realistic
string $u$ from only the overlap graph of $u$. In this way we can
define the reduction graph for realistic overlap graphs in a direct
way.

Next we consider sets of overlapping pointers corresponding to pairs
of vertices in reduction graphs, and start to develop a calculus for
these sets that will later enable us to characterize the existence
of certain edges in the reduction graph, cf.
Theorem~\ref{overlap_to_redgraph}.
\begin{Example}
We again consider the legal string $u = 3 2 \bar 4 3 \bar 2 4$ and
its reduction graph $\fredgr_u$ from
Example~\ref{ex_red_graph_of_legal_string}. Desire edge $e =
\{\RGVertR{2}, \RGVertR{5}\}$ is connected to reality edges $e_1 =
\{\RGVertR{2}, \RGVertL{3}\}$ and $e_2 = \{\RGVertR{5},
\RGVertL{6}\}$ with positions $2$ and $5$ respectively. We have
$O_u(2,5) = \{2,3,4\}$. Also, reality edges $\{\RGVertR{1},
\RGVertL{2}\}$ and $\{\RGVertR{2}, \RGVertL{3}\}$ have positions $1$
and $2$ respectively. We have $O_u(1,2) = \{2\}$.
\end{Example}

\begin{Lemma} \label{overlap_edge_1}
Let $u$ be a legal string. Let $e = \{v_1,v_2\}$ be a desire edge of
$\fredgr_u$ and let $p$ be the label of both $v_1$ and $v_2$. Then
$$
O_u(\posn(v_1),\posn(v_2)) = \begin{cases} O_u(p) & \mbox{if $p$ is negative in $u$},\\
O_u(p) \oplus \{p\} & \mbox{if $p$ is positive in $u$}. \end{cases}
$$
\end{Lemma}
%
%
%
\begin{Proof}
Let $u = p_1 p_2 \ldots p_n$ with $p_1, p_2, \ldots, p_n \in \Pi$
and let $i$ and $j$ be such that $i<j$ and $p = p_i = p_j$. Without
loss of generality, we can assume $\posn(v_1) < \posn(v_2)$. Then,
$v_1 \in \{\RGVertL{i}, \RGVertR{i}\}$ and $v_2 \in \{\RGVertL{j},
\RGVertR{j}\}$, hence $\posn(v_1) \in \{i-1,i\}$ and $\posn(v_2) \in
\{j-1,j\}$.

First, assume that $p$ is negative in $u$. By the definition of
reduction graph, the following two cases are possible:
\begin{enumerate}
\item $e = \{ \RGVertL{i}, \RGVertR{j} \}$, thus
$O_u(\posn(\RGVertL{i}),\posn(\RGVertR{j})) = O_u(i-1,j) = O_u(p)$,
\item $e = \{ \RGVertR{i}, \RGVertL{j} \}$, thus $O_u(\posn(\RGVertL{i}),\posn(\RGVertR{j-1})) = O_u(i,j-1) = O_u(p)$,
\end{enumerate}
Thus in both cases we have $O_u(\posn(v_1),\posn(v_2)) = O_u(p)$.

Finally, assume that $p$ is positive in $u$. By the definition of
reduction graph, the following two cases are possible:
\begin{enumerate}
\item $e = \{ \RGVertL{i}, \RGVertL{j} \}$, thus
$O_u(\posn(\RGVertL{i}),\posn(\RGVertL{j})) = O_u(i-1,j-1) = O_u(p)
\oplus \{p\}$,
\item $e = \{ \RGVertR{i}, \RGVertR{j} \}$, thus $O_u(\posn(\RGVertR{i}),\posn(\RGVertR{j})) = O_u(i,j) = O_u(p) \oplus \{p\}$,
\end{enumerate}
Thus in both cases we have $O_u(i_1,i_2) = O_u(p) \oplus \{p\}$.
\end{Proof}

The following result follows by iteratively applying the previous
lemma.
\begin{Corollary} \label{overlap_edge_1_iterative}
Let $u$ be a legal string. Let
$$
\xymatrix @=18pt{ p_0 \redge[r] & p_1 \dedge[r] & p_1 \redge[r] &
p_2 \dedge[r] & p_2 \redge[r] & .. \redge[r] & p_n \dedge[r] & p_n
\redge[r] & p_{n+1}}
$$
be a subgraph of $\fredgr_u$, where (as usual) the vertices in the
figure are represented by their labels, and let $e_1$ ($e_2$, resp.)
be the leftmost (rightmost, resp.) edge. Note that $e_1$ and $e_2$
are reality edges and therefore $\posn(e_1)$ and $\posn(e_2)$ are
defined. Then $O_u(\posn(e_1),\posn(e_2)) = \left( \posve(u) \cap P
\right) \oplus \left(\bigoplus_{t \in P} O_u(t)\right)$ with $P =
\{p_1,\ldots,p_n\}$.
\end{Corollary}

By the definition of the reduction graph the following lemma holds.
\begin{Lemma} \label{overlap_edge_2}
Let $u$ be a legal string. If $\RGVertL{i}$ and $\RGVertR{i}$ are
vertices of $\fredgr_u$, then
$O_u(\posn(\RGVertL{i}),\posn(\RGVertR{i})) = \{p\}$, where $p$ is
the label of $\RGVertL{i}$ and $\RGVertR{i}$.
\end{Lemma}

\begin{Example}
We again consider the legal string $u$ and desire edge $e$ as in the
previous example. Since $e$ has vertices labelled by positive
pointer $2$, by Lemma~\ref{overlap_edge_1} we have (again) $O_u(2,5)
= O_u(2) \oplus \{2\} = \{2,3,4\}$. Also, since $\RGVertL{2}$ and
$\RGVertR{2}$ with positions $1$ and $2$ respectively are labelled
by $2$, by Lemma~\ref{overlap_edge_2} we have (again) $O_u(1,2) =
\{2\}$.
\end{Example}

\section{The Reduction Graph of Realistic Strings}
\label{sect_reduction_graph_realistic} The next theorem asserts that
overlap graph $\overlapgr$ for realistic string $u$ retains all
information of $\fredgr_u$ (up to isomorphism). In the next few
sections, we will give a method to determine $\fredgr_u$ (up to
isomorphism), given $\overlapgr$. Of course, the naive method is to
first determine a legal string $u$ corresponding to $\overlapgr$ and
then to determine the reduction graph of $u$. However, we present a
method that is able to construct $\fredgr_u$ in a direct way from
$\overlapgr$.

\begin{Theorem} \label{one_to_one_overlap_redgraph}
Let $u$ and $v$ be realistic strings. If $\overlapgru{u} =
\overlapgru{v}$, then $\fredgr_u \approx \fredgr_v$.
\end{Theorem}
\begin{Proof}
By Theorem~1 in \cite{DBLP:conf/birthday/HarjuPR04} (or Theorem~10.2
in \cite{GeneAssemblyBook}), we have $\overlapgru{u} =
\overlapgru{v}$ iff $v$ can be obtained from $u$ by a composition of
reversal, complement and conjugation operations. By the definition
of reduction graph it is clear that the reduction graph is invariant
under these operations (up to isomorphism). Thus, $\fredgr_u \approx
\fredgr_v$.
\end{Proof}

\begin{figure}
$$
\xymatrix @=15pt{
2^- \ar@{-}[dr] &   & 4^- \ar@{-}[dl] \\
  & 3^- & \\
6^- \ar@{-}[ur] &   & 5^- \ar@{-}[ul]
}
$$
\caption{The overlap graph of both legal strings $u$ and $v$ of
Example~\ref{ex_legal_not_1_to_1_overlap_redgraph}.}
\label{ex_legal_not_1_to_1_overlap_redgraph_fig1}
\end{figure}
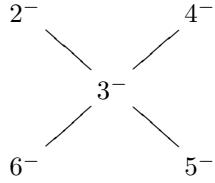

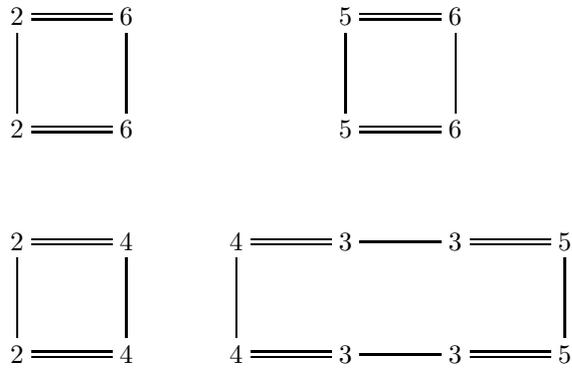
\begin{figure}
$$
\xymatrix @=30pt{
& 2 \redge[r] \dedge[d] & 6 \dedge[d] & & 5 \dedge[d] \redge[r] & 6 \dedge[d]\\
& 2 \redge[r] & 6 & & 5 \redge[r] & 6\\
& 2 \redge[r] \dedge[d] & 4 \dedge[d] & 4 \dedge[d] \redge[r] & 3 \dedge[r] & 3 \redge[r] & 5 \dedge[d] \\
& 2 \redge[r] & 4 & 4 \redge[r] & 3 \dedge[r] & 3 \redge[r] & 5
}
$$
\caption{The reduction graph of $u$ of
Example~\ref{ex_legal_not_1_to_1_overlap_redgraph}.}
\label{ex_legal_not_1_to_1_overlap_redgraph_fig2}
\end{figure}

The previous theorem is \emph{not} true for legal strings in general
--- the next two examples illustrate that
legal strings having the same overlap graph can have different
reduction graphs.
\begin{Example} \label{ex_legal_not_1_to_1_overlap_redgraph}
Let $u = 2653562434$ and $v = h(u)$, where $h$ is the homomorphism
that interchanges $5$ and $6$.
%
%
Thus, $v = 2563652434$. Note that both $u$ and $v$ are not
realistic, because substrings $535$ of $u$ and $636$ of $v$ can
obviously not be substrings of realistic strings. The overlap graph
of $u$ is depicted in
Figure~\ref{ex_legal_not_1_to_1_overlap_redgraph_fig1}. From
Figure~\ref{ex_legal_not_1_to_1_overlap_redgraph_fig1} and the fact
that $v$ is obtained from $u$ by renumbering $5$ and $6$, it follows
that the overlap graphs of $u$ and $v$ are equal. The reduction
graph $\fredgr_u$ of $u$ is depicted in
Figure~\ref{ex_legal_not_1_to_1_overlap_redgraph_fig2}. The
reduction graph $\fredgr_v$ of $v$ is obtained from $\fredgr_u$ by
renumbering the labels of the vertices according to $h$. Clearly,
$\fredgr_u \not\approx \fredgr_v$.
\end{Example}

\begin{figure}
$$
\xymatrix @=30pt{
2 \redge[d] \dedge@/^1.0pc/[d] & 3 \redge[d] \dedge@/^1.0pc/[d] & 4 \redge[d] \dedge@/^1.0pc/[d] & 2 \dedge[r] \redge[d] & 2 \redge[r] & 3 \dedge[d]\\
2 & 3 & 4 & 4 \dedge[r] & 4 \redge[r] & 3}
$$
\caption{The reduction graph of $u$ of
Example~\ref{ex2_legal_not_1_to_1_overlap_redgraph}.}
\label{ex2_legal_not_1_to_1_overlap_redgraph_fig1}
\end{figure}
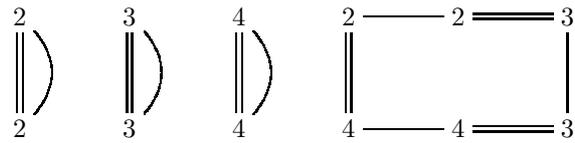

\begin{figure}
$$
\xymatrix @=30pt{
2 \redge[d] \dedge@/^1.0pc/[d] & 4 \redge[d] \dedge@/^1.0pc/[d] & 2 \dedge[d] \redge[r] & 3 \dedge[d] & 3 \dedge[d] \redge[r] & 4 \dedge[d]\\
2 & 4 & 2 \redge[r] & 3 & 3 \redge[r] & 4}
$$
\caption{The reduction graph of $v$ of
Example~\ref{ex2_legal_not_1_to_1_overlap_redgraph}.}
\label{ex2_legal_not_1_to_1_overlap_redgraph_fig2}
\end{figure}
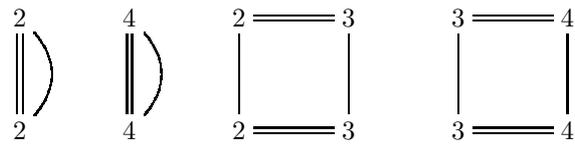

\begin{Example} \label{ex2_legal_not_1_to_1_overlap_redgraph}
Let $u = \pi_{\kappa}(M_1 M_2 M_3 M_4) = 223344$ be a realistic
string and let $v = 234432$ be a legal string. Note that $v$ is not
realistic. Legal strings $u$ and $v$ have the same overlap graph
$\overlapgr$ ($\overlapgr = (\{2,3,4\},\emptyset,\sigma,\{+,-\})$,
where $\sigma(v) = -$ for $v \in \{2,3,4\}$). The reduction graph
$\fredgr_u$ of $u$ is depicted in
Figure~\ref{ex2_legal_not_1_to_1_overlap_redgraph_fig1}, and the
reduction graph $\fredgr_v$ of $v$ is depicted in
Figure~\ref{ex2_legal_not_1_to_1_overlap_redgraph_fig2}. Note that
$\fredgr_u$ has a component consisting of six vertices, while
$\fredgr_v$ does not have such a component. Therefore, $\fredgr_u
\not\approx \fredgr_v$.
\end{Example}

For realistic strings the reduction graph has a special form. This
is seen as follows. For $1 < i < \kappa$ the symbol $M_i$ (or
$\bar{M}_i$) in the micronuclear arrangement defines two pointers
$p_i$ and $p_{i+1}$ (or $\bar{p}_{i+1}$ and $\bar{p}_{i}$) in the
corresponding realistic string $u$. At the same time the substring
$p_i p_{i+1}$ (or $\bar{p}_{i+1} \bar{p}_{i}$, resp.) of $u$
corresponding to $M_i$ (or $\bar{M}_i$, resp.) defines four vertices
$\RGVertL{j}, \RGVertR{j}, \RGVertL{j+1}, \RGVertR{j+1}$ in
$\fredgr_u$. It is easily verified (cf.
Theorem~\ref{micronuclear_to_root_subgraph} below) that the
``middle'' two vertices $\RGVertR{j}$ and $\RGVertL{j+1}$, labelled
by $p_i$ and $p_{i+1}$ respectively, are connected by a reality edge
and $\RGVertR{j}$ ($\RGVertL{j+1}$, resp.) is connected by a desire
edge to a ``middle vertex'' resulting from $M_{i-1}$ or
$\bar{M}_{i-1}$ ($M_{i+1}$ or $\bar{M}_{i+1}$, resp.). This leads to
the following definition.
\begin{Definition}
Let $u$ be a legal string and let $\kappa = |\dom(u)|+1$. If
$\fredgr_u$ contains a subgraph $L$ of the following form:
$$
\xymatrix @=18pt{ 2 \dedge[r] & 2 \redge[r] & 3 \dedge[r] & 3
\redge[r] & .. \redge[r] & \kappa \dedge[r] & \kappa }
$$
where the vertices in the figure are represented by their labels,
then we say that $u$ is \emph{rooted}. Subgraph $L$ is called a
\emph{root subgraph of $\fredgr_u$}.
\end{Definition}

\begin{Example}
The realistic string $u$ with $\dom(u) = \{2,3,\ldots,7\}$ in
Example~\ref{ex_red_graph_of_realistic_string} is rooted because the
reduction graph of $u$, depicted in
Figure~\ref{ex_red_graph_of_realistic_string_fig1}, contains the
subgraph
$$
\xymatrix @=18pt{ 2 \dedge[r] & 2 \redge[r] & 3 \dedge[r] & 3
\redge[r] & .. \redge[r] & 7 \dedge[r] & 7 }
$$
\end{Example}
The next theorem shows that indeed every realistic string is rooted.

\begin{Theorem} \label{micronuclear_to_root_subgraph}
Every realistic string is rooted.
\end{Theorem}
\begin{Proof}
Consider a micronuclear arrangement for a realistic string $u$. Let
$\kappa = |\dom(u)|+1$. By the definition of $\pi_{\kappa}$, there
is a reality edge $e_i$ (corresponding to either $\pi_{\kappa}(M_i)
= i(i+1)$ or $\pi_{\kappa}(\overline{M_i}) = \overline{(i+1)} \
\overline{i}$) connecting a vertex labelled by $i$ to a vertex
labelled by $i+1$ for each $2 \leq i < \kappa$. It suffices to prove
that there is a desire edge connecting $e_i$ to $e_{i+1}$ for each
$2 \leq i < \kappa - 1$. This can easily be seen by checking the
four cases where $e_i$ corresponds to either $\pi_{\kappa}(M_i)$ or
$\pi_{\kappa}(\overline{M_i})$ and $e_{i+1}$ corresponds to either
$\pi_{\kappa}(M_{i+1})$ or $\pi_{\kappa}(\overline{M_{i+1}})$.
\end{Proof}
In the remaining of this paper, we will denote $|\dom(u)|+1$ by
$\kappa$ for rooted strings, when it is clear which rooted string
$u$ is meant. The reduction graph of a realistic string may have
more than one root subgraph: it is easy to verify that realistic
string $2 3 4 \cdots \kappa 2 3 4 \cdots \kappa$ for $\kappa \geq 2$
has two root subgraphs.

Example~\ref{ex_red_graph_of_legal_string} shows that not every
rooted string is realistic. The remaining results that consider
realistic strings also hold for rooted strings, since we will not be
using any properties of realistic string that are not true for
rooted strings in general.

For a given root subgraph $L$, it is convenient to uniquely identify
every reality edge containing a vertex of $L$. This is done through
the following definition.

\begin{Definition}
Let $u$ be a rooted string and let $L$ be a root subgraph of
$\fredgr_u$. We define $\rspos_{L,k}$ for $2 \leq k < \kappa$ as the
position of the edge of $L$ that has vertices labelled by $k$ and
$k+1$. We define $\rspos_{L,1}$ ($\rspos_{L,\kappa}$, resp.) as the
position of the edge of $\fredgr_u$ not in $L$ containing a vertex
of $L$ labelled by $2$ ($\kappa$, resp.). When $\kappa = 2$, to
ensure that $\rspos_{L,1}$ and $\rspos_{L,\kappa}$ are well defined,
we additionally require that $\rspos_{L,1} < \rspos_{L,\kappa}$.
\end{Definition}
Thus, $\rspos_{L,k}$ (for $1 \leq k \leq \kappa$) uniquely
identifies every reality edge containing a vertex of $L$. If it is
clear which root subgraph $L$ is meant, we simply write $\rspos_{k}$
instead of $\rspos_{L,k}$ for $1 \leq k \leq \kappa$.

The next lemma is essential to prove the main theorem of this paper.
\begin{Lemma} \label{overlap_equal_pos}
Let $u$ be a rooted string. Let $L$ be a root subgraph of
$\fredgr_u$. Let $i$ and $j$ be positions of reality edges in
$\fredgr_u$ that are not edges of $L$. Then $O_u(i,j) = \emptyset$
iff $i=j$.
\end{Lemma}
\begin{Proof}
The reverse implication is trivially satisfied. We now prove the
forward implication. The reality edge $e_k$ (for $2 \leq k <
\kappa$) in $L$ with vertices labelled by $k$ and $k+1$ corresponds
to a cyclic substring $\tilde{M}_k \in \{p_1 p_2, p_2 p_1 \mid p_1
\in \{k, \overline{k}\}, p_2 \in \{k+1, \overline{k+1}\}\}$ of $u$.
Let $k_1$ and $k_2$ with $2 \leq k_1 < k_2 < \kappa$. If $k_1 + 1 =
k_2$, then $e_{k_1}$ and $e_{k_2}$ are connected by a desire edge
(by the definition of $L$). Therefore, pointer $k_2$ common in
$\tilde{M}_{k_1}$ and $\tilde{M}_{k_2}$ originates from two
different occurrences in $u$. If on the other hand $k_1 + 1 \not=
k_2$, then $\tilde{M}_{k_1}$ and $\tilde{M}_{k_2}$ do not have a
letter in common. Therefore, in both cases, $\tilde{M}_{k_1}$ and
$\tilde{M}_{k_2}$ are disjoint cyclic substrings of $u$. Thus the
$\tilde{M}_k$ for $2 \leq k < \kappa$ are pairwise disjoint cyclic
substrings of $u$.

Without loss of generality assume $i \leq j$. Let $u = u_1 u_2
\cdots u_n$ with $u_i \in \Pi$. Since $u$ is a legal string, every
$u_{l}$ for $1 \leq l \leq n$ is either part of a $\tilde{M}_k$
(with $2 \leq k < \kappa$) or in $\{2, \bar 2, \kappa, \bar
\kappa\}$. Consider $u' = u_{i+1} u_{i+2} \cdots u_{j}$. Since $i$
and $j$ are positions of reality edges in $\fredgr_u$ that are not
edges of $L$, we have $u' = \tilde{M}_{k_1} \tilde{M}_{k_2} \cdots
\tilde{M}_{k_m}$ for some distinct $k_1, k_2, \ldots, k_{m} \in
\{1,2,\ldots,\kappa\}$, where $\tilde{M}_1 \in \{2, \bar 2\}$ and
$\tilde{M}_{\kappa} \in \{\kappa, \bar \kappa\}$.

It suffices to prove that $u' = \lambda$. Assume to the contrary
that $u' \not= \lambda$. Then there is a $1 \leq l \leq \kappa$ such
that $\tilde{M}_l$ is a substring of $u'$. Because $O_u(i,j) =
\emptyset$, we know that $u'$ is legal. If $l
> 1$, then $\tilde{M}_{l-1}$ is also a substring of $u'$, otherwise
$u'$ would not be a legal string. Similarly, if $l < \kappa$, then
$\tilde{M}_{l+1}$ is also a substring of $u'$. By iteration, we
conclude that $u' = u$. Therefore, $i = 0$. This is a contradiction,
since $0$ cannot be a position of a reality edge. Thus, $u' =
\lambda$.
\end{Proof}

\begin{Lemma} \label{overlap_edge_realistic}
Let $u$ be a rooted string. Let $L$ be a root subgraph of
$\fredgr_u$. If $\RGVertL{i}$ and $\RGVertR{i}$ are vertices of
$\fredgr_u$, then exactly one of $\RGVertL{i}$ and $\RGVertR{i}$
belongs to $L$.
\end{Lemma}
\begin{Proof}
By the definition of reduction graph, $\RGVertL{i}$ and
$\RGVertR{i}$ have a common vertex label $p$ but are not connected
by a desire edge. Therefore, $\RGVertL{i}$ and $\RGVertR{i}$ do not
both belong to $L$. Now, if $\RGVertL{i}$ and $\RGVertR{i}$ both do
not belong to $L$, then the other vertices labelled by $p$, which
are $\RGVertL{j}$ and $\RGVertR{j}$ for some $j$, both belong to $L$
-- a contradiction by the previous argument. Therefore, either
$\RGVertL{i}$ or $\RGVertR{i}$ belongs to $L$, and the other one
does not belong to $L$.
\end{Proof}

The next result provides the main idea to determine the reduction
graph given (only) the overlap graph as presented in
Section~\ref{sect_overlap_to_red_graph}. It relies heavily on the
previous lemmas.
\begin{Theorem} \label{th_partone_char}
Let $u$ be a rooted string, let $L$ be a root subgraph of
$\fredgr_u$, and let $p,q \in \dom(u)$ with $p < q$. Then there is a
reality edge $e$ in $\fredgr_u$ with both vertices not in $L$, one
vertex labelled by $p$ and the other labelled by $q$ iff
$$ \bigoplus_{t \in P} O_{u}(t) = \left( \posve(u) \cap P \right)
\oplus \{p\} \oplus \{q\},
$$
where $P = \{p+1,\ldots,q-1\} \cup P'$ for some $P' \subseteq
\{p,q\}$.
\end{Theorem}
\begin{Proof}
We first prove the forward implication. Let $e = \{v_1,v_2\}$ with
$v_1$ labelled by $p$, $v_2$ labelled by $q$, and $\posn(e) = i$.
Thus $e = \{ \RGVertR{i}, \RGVertL{i+1} \}$. We assume that $v_1 =
\RGVertR{i}$ and $v_2 = \RGVertL{i+1}$, the other case is proved
similarly. Let $i_1 = \posn(\RGVertL{i})$ and $i_2 =
\posn(\RGVertR{i+1})$. By Lemma~\ref{overlap_edge_2} $O_u(i,i_1) =
\{p\}$ and $O_u(i_2,i) = \{q\}$. By
Lemma~\ref{overlap_edge_realistic}, $\RGVertL{i}$ (labelled by $p$)
and $\RGVertR{i+1}$ (labelled by $q$) belong to $L$. Thus $i_1 \in
\{ \rspos_{p-1}, \rspos_p \}$ and $i_2 \in \{ \rspos_{q-1}, \rspos_q
\}$. By applying Corollary~\ref{overlap_edge_1_iterative} on $L$, we
have $O_u(i_1,i_2) = \left( \posve(u) \cap P \right) \oplus
\left(\bigoplus_{t \in P} O_u(t)\right)$ with $P =
\{p+1,\ldots,q-1\} \cup P'$ for some $P' \subseteq \{p,q\}$. By
definition of $O_u(i,j)$ we have
$$
\emptyset = O_u(i,i) = O_u(i,i_1) \oplus O_u(i_1,i_2) \oplus
O_u(i_2,i)
$$
Thus the desired result follows.

We now prove the reverse implication. By applying
Corollary~\ref{overlap_edge_1_iterative} on $L$, we have
$O_u(i_1,i_2) = \left( \posve(u) \cap P \right) \oplus
\left(\bigoplus_{t \in P} O_u(t)\right)$ for some $i_1 \in \{
\rspos_{p-1}, \rspos_p \}$ and $i_2 \in \{ \rspos_{q-1}, \rspos_q
\}$ (depending on $P'$). By Lemma~\ref{overlap_edge_2}, there is a
vertex $v_1$ ($v_2$, resp.) labelled by $p$ ($q$, resp.) with
position $i$ ($j$, resp.) such that $O_u(i,i_1) = \{p\}$ and
$O_u(i_2,j) = \{q\}$. By Lemma~\ref{overlap_edge_realistic} these
vertices are not in $L$. We have now
$$
\emptyset = O_u(i,i_1) \oplus O_u(i_1,i_2) \oplus O_u(i_2,j) =
O_u(i,j)
$$
By Lemma~\ref{overlap_equal_pos}, $O_u(i,j) = \emptyset$ implies
that $i=j$. Thus, there is a reality edge $\{v_1,v_2\}$ in
$\fredgr_u$ (with position $i$), such that $v_1$ is labelled by $p$
and $v_2$ is labelled by $q$ and both are not vertices of $L$.
\end{Proof}

Let $\overlapgr_u$ be the overlap graph of some legal string $u$.
Clearly we have $\posve(u) = \posve(\overlapgr_u)$ and for all $p
\in \dom(u) = \dom(\overlapgr_u)$, $O_{u}(p) = O_{\overlapgr_u}(p)$.
Thus by Theorem~\ref{th_partone_char} we can determine, given the
overlap graph of a rooted string $u$, if there is a reality edge in
$\fredgr_u$ with both vertices outside $L$ that connects a vertex
labelled by $p$ to a vertex labelled by $q$. We will extend this
result to completely determine the reduction graph given the overlap
graph of a rooted string (or a realistic string in particular).

\section{Compressing the Reduction Graph} \label{sect_compress_function}
In this section we define the $\compress$ function. The $\compress$
function simplifies reduction graphs by replacing the subgraph $
\xymatrix @=18pt{ p \dedge[r] & p} $ by a single vertex labelled by
$p$. In this way, one can simplify reduction graphs without ``losing
information''. We will define $\compress$ for a general family of
graphs $\mathcal{G}$ which includes all reduction graphs. The formal
definitions of $\mathcal{G}$ and $\compress$ are given below.

Let $\mathcal{G}$ be the set of 2-edge coloured graphs $G = (V, E_1,
E_2, f, \Gamma)$ with the property that for all $\{v_1, v_2\} \in
E_2$, it holds that $f(v_1) = f(v_2)$. Note that for a reduction
graph $\fredgr_u$, we have $\fredgr_u \in \mathcal{G}$ because both
vertices of a desire edge have the same label. For all $G \in
\mathcal{G}$, $\compress(G)$ is obtained from $G$ by considering the
second set of edges as vertices in the labelled graph. Thus, for the
case when $G$ is a reduction graph, the function $\compress$
``compresses'' the desire edges to vertices.
\begin{Definition}
The function $\compress$ from $\mathcal{G}$ to the set of labelled
graphs is defined as follows. Let $G = (V, E_1, E_2, f, \Gamma) \in
\mathcal{G}$, then
$$
\compress(G) = (E_2, E'_1, f', \Gamma)
$$
is a labelled graph, where
$$
E'_1 = \{ \{e_1,e_2\} \subseteq E_2 \mid \exists v_1, v_2 \in V: v_1
\in e_1, v_2 \in e_2, e_1 \not= e_2 \mbox{ and } \{v_1, v_2\} \in
E_1 \},
$$
and for $e \in E_2$: $f'(e) = f(v)$ with $v \in e$.
\end{Definition}
Note that $f'$ is well defined, because for all $\{v_1, v_2\} \in
E_2$, it holds that $f(v_1) = f(v_2)$.

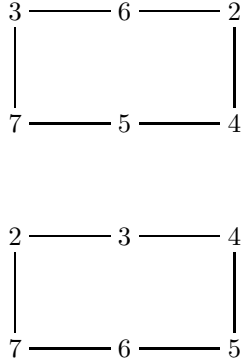
\begin{figure}
$$
\xymatrix @=30pt{
3 \ar@{-}[r] & 6 \ar@{-}[r] & 2 \ar@{-}[d] \\
7 \ar@{-}[u] & 5 \ar@{-}[l] & 4 \ar@{-}[l] \\
2 \ar@{-}[r] & 3 \ar@{-}[r] & 4 \ar@{-}[d] \\
7 \ar@{-}[u] & 6 \ar@{-}[l] & 5 \ar@{-}[l] }
$$
\caption{The labelled graph $\compress(\fredgr_u)$, where
$\fredgr_u$ is defined in Example~\ref{ex_compress_realistic_graph}.
The vertices in the figure are represented by their labels.}
\label{ex_compress_realistic_graph_fig1}
\end{figure}

\begin{Example} \label{ex_compress_realistic_graph}
We are again considering the realistic string $u$ defined in
Example~\ref{ex_red_graph_of_realistic_string}. The reduction graph
of $\fredgr_u$ is depicted in
Figure~\ref{ex_red_graph_of_realistic_string_fig1}. The labelled
graph $\compress(\fredgr_u)$ is depicted in
Figure~\ref{ex_compress_realistic_graph_fig1}. Since this graph has
just one set of edges, the reality edges are depicted as `single
edges' instead of `double edges' as we did for reduction graphs.
\end{Example}

It is not hard to see that for reduction graphs $\fredgr_u$ and
$\fredgr_v$, we have $\fredgr_u \approx \fredgr_v$ iff
$\compress(\fredgr_u) \approx \compress(\fredgr_v)$. In this sense,
function $\compress$ allows one to simplify reduction graphs without
losing information.

\section{From Overlap Graph to Reduction Graph}
\label{sect_overlap_to_red_graph} Here we define reduction graphs
for realistic overlap graphs, inspired by the characterization of
Theorem~\ref{th_partone_char}. In the remaining part of this section
we will show its equivalence with reduction graphs for realistic
strings.
\begin{Definition} \label{def_reduction_graph_overlap_graph}
Let $\overlapgr = (Dom_{\overlapgr},E_{\overlapgr},\sigma,\{+,-\})$
be a realistic overlap graph and let $\kappa = |Dom_{\gamma}|+1$.
The \emph{reduction graph of $\overlapgr$}, denoted by
$\fredgr_{\overlapgr}$, is a labelled graph
$$
\fredgr_{\overlapgr} = (V,E,f,Dom_{\overlapgr}),
$$
where
$$
V = \{ \RGOVertNonRoot{p}, \RGOVertRoot{p} \mid 2 \leq p \leq \kappa
\},
$$
$$
\mbox{$f(\RGOVertNonRoot{p}) = f(\RGOVertRoot{p}) = p$, for $2 \leq
p \leq \kappa$, and}
$$
$e \in E$ iff one of the following conditions hold:
\begin{enumerate}
\item $e = \{\RGOVertRoot{p},\RGOVertRoot{p+1}\}$ and $2 \leq p <
\kappa$.
\item $e = \{\RGOVertNonRoot{p},\RGOVertNonRoot{q}\}$, $2 \leq p < q \leq \kappa$, and
$$
\bigoplus_{t \in P} O_{\overlapgr}(t) = \left( \posve(\overlapgr)
\cap P \right) \oplus \{p\} \oplus \{q\},
$$
where $P = \{p+1,\ldots,q-1\} \cup P'$ for some $P' \subseteq
\{p,q\}$.
\item $e = \{\RGOVertRoot{2},\RGOVertNonRoot{p}\}$, $2 \leq p \leq \kappa$, and
$$
\bigoplus_{t \in P} O_{\overlapgr}(t) = \left( \posve(\overlapgr)
\cap P \right) \oplus \{p\},
$$
where $P = \{2,\ldots,p-1\} \cup P'$ for some $P' \subseteq \{p\}$.
\item $e = \{\RGOVertRoot{\kappa},\RGOVertNonRoot{p}\}$, $2 \leq p \leq \kappa$, and
$$
\bigoplus_{t \in P} O_{\overlapgr}(t) = \left( \posve(\overlapgr)
\cap P \right) \oplus \{p\},
$$
where $P = \{p+1,\ldots,\kappa\} \cup P'$ for some $P' \subseteq
\{p\}$.
\item $e = \{\RGOVertRoot{2},\RGOVertRoot{\kappa}\}$, $\kappa > 3$, and
$$
\bigoplus_{t \in P} O_{\overlapgr}(t) = \posve(\overlapgr) \cap P,
$$
where $P = \{2,\ldots,\kappa\}$.
\end{enumerate}
\end{Definition}

\begin{figure}
$$
\xymatrix @=30pt{
2^- \ar@{-}[r] & 3^- \ar@{-}[r] \ar@{-}[d] \ar@/^0.0pc/@{-}[dr] \ar@/^0.0pc/@{-}[dl] & 4^- \ar@{-}[d] \\
6^- & 7^- \ar@{-}[r] & 5^- }
$$
\caption{The overlap graph $\overlapgr$ of a realistic string (used
in Example~\ref{ex1_overlap_graph}).} \label{ex1_fig1_overlap_graph}
\end{figure}

\def\pijlrc#1{\ar@/^0.3pc/@{-}[r]}
\def\pijllc#1{\ar@/^0.3pc/@{-}[l]}
\def\pijlr#1{\ar@{-}[r]}
\def\pijll#1{\ar@{-}[l]}
\def\pijld#1{\ar@{-}[d]}
\def\pijlu#1{\ar@{-}[u]}
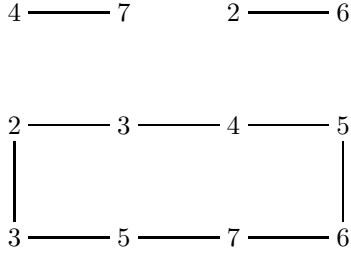
\begin{figure}
$$
\xymatrix @=30pt{
4 \pijlr{7} & 7 & 2 \pijlr{3,5,6} & 6 \\
2 \pijlr{2,5,6} & 3 \pijlr{4,7} & 4 \pijlr{3,4,5,7} & 5 \pijld{5} \\
3 \pijlu{2,3,5,6} & 5 \pijll{3,4,7} & 7 \pijll{} & 6 \pijll{3,5} }
$$
\caption{The reduction graph $\fredgr_\overlapgr$ of the overlap
graph $\overlapgr$ of Example~\ref{ex1_overlap_graph}. The vertices
in the figure are represented by their labels.}
\label{ex1_fig2_red_graph}
\end{figure}

\begin{Example} \label{ex1_overlap_graph}
The overlap graph $\overlapgr$ in
Figure~\ref{ex1_fig1_overlap_graph} is realistic. Indeed, for
example realistic string $u = \pi_7 (M_4 M_3 M_7 M_5 M_2 M_1 M_6) =
453475623267$ has this overlap graph. Clearly, the reduction graph
$\fredgr_\overlapgr$ of $\overlapgr$ has the edges
$\{\RGOVertRoot{p},\RGOVertRoot{p+1}\}$ for $2 \leq p < 7$. The
following table lists the remaining edges of $\fredgr_\overlapgr$.
The table also states the characterizing conditions for each edge as
stated in Definition~\ref{def_reduction_graph_overlap_graph}. Note
that $\posve(\overlapgr) = \emptyset$, and consequently the
right-hand side of the defining equations in points 2, 3 and 4 in
Definition~\ref{def_reduction_graph_overlap_graph} are independent
of the choice of $P'$.

\begin{tabular}[t]{c|c|r}
Edge & $P$ & Witness \\
\hline $\{J_2,J_6\}$ & $\{3,4,5\}$ & $\{2,4,5,6,7\} \oplus \{3,5\}
\oplus \{3,4,7\} = \{2,6\}$ \\
$\{J_2,J_6\}$ & $\{2,3,4,5,6\}$ & $\{3\} \oplus \{2,4,5,6,7\} \oplus
\{3,5\} \oplus \{3,4,7\} \oplus \{3\} = \{2,6\}$ \\
$\{J_4,J_7\}$ & $\{5,6\}$ & $\{3,4,7\} \oplus \{3\}= \{4,7\}$ \\
$\{J_4,J_7\}$ & $\{4,5,6,7\}$ & $\{3,5\} \oplus \{3,4,7\} \oplus \{3\} \oplus \{3,5\} = \{4,7\}$ \\
$\{J_3,J_5\}$ & $\{4\}$ & $\{3,5\} = \{3,5\}$\\
$\{J_5,J'_7\}$ & $\{6,7\}$ & $\{3\} \oplus \{3,5\} = \{5\}$ \\
$\{J'_2,J_3\}$ & $\{2\}$ & $\{3\} = \{3\}$
\end{tabular}

We have now completely determined $\fredgr_\overlapgr$; it is shown
in Figure~\ref{ex1_fig2_red_graph}. As we have done for reduction
graphs of legal strings, in the figures, the vertices of reduction
graphs of realistic overlap graphs are represented by their labels.
\end{Example}

\begin{figure}
$$
\xymatrix @=30pt{ & 2^+ \ar@{-}[dl] \ar@{-}[dr] \ar@{-}[drr] \\
5^- \ar@{-}[rr] \ar@{-}[dr] \ar@{-}[ddr] & & 4^- \ar@{-}[dl] \ar@{-}[ddl] & 7^- \ar@{-}[ddll]\\
& 3^+ \ar@{-}[d] \\
& 6^- }
$$
\caption{The overlap graph $\overlapgr$ of a realistic string (used
in Example~\ref{ex2_overlap_graph}).} \label{ex2_fig1_overlap_graph}
\end{figure}

\begin{figure}
$$
\xymatrix @=30pt{
3 \ar@{-}[r] & 6 \ar@{-}[r] & 2 \ar@{-}[d] \\
7 \ar@{-}[u] & 5 \ar@{-}[l] & 4 \ar@{-}[l] \\
2 \ar@{-}[r] & 3 \ar@{-}[r] & 4 \ar@{-}[d] \\
7 \ar@{-}[u] & 6 \ar@{-}[l] & 5 \ar@{-}[l] }
$$
\caption{The reduction graph $\fredgr_\overlapgr$ of the overlap
graph $\overlapgr$ of Example~\ref{ex2_overlap_graph}.}
\label{ex2_fig2_red_graph}
\end{figure}

\begin{Example} \label{ex2_overlap_graph}
In the second example we construct the reduction graph of an overlap
graph that contains positive pointers. The overlap graph
$\overlapgr$ in Figure~\ref{ex2_fig1_overlap_graph} is realistic.
Indeed, for example realistic string $u = \pi_7(M_7 M_1 M_6 M_3 M_5
\overline{M_2} M_4)= 72673456 \bar 3 \bar 2 45$ introduced in
Example~\ref{ex_red_graph_of_realistic_string} has this overlap
graph. Again, the reduction graph $\fredgr_\overlapgr$ of
$\overlapgr$ has the edges $\{\RGOVertRoot{p},\RGOVertRoot{p+1}\}$
for $2 \leq p < 7$. The remaining edges are listed in the table
below.

\begin{tabular}[t]{c|c|r}
Edge & $P$ & Witness \\
\hline $\{J_3,J_7\}$ & $\{4,5,6\}$ & $\{2,3,5,6\} \oplus \{2,3,4,6\}
\oplus \{3,4,5,7\} = \{3,7\}$ \\
$\{J_3,J_6\}$ & $\{3,4,5\}$ & $\{3\} \oplus \{4,5,6\} \oplus
\{2,3,5,6\} \oplus \{2,3,4,6\} = \{3,6\}$ \\
$\{J_2,J_6\}$ & $\{2,3,4,5,6\}$ & $\{2\} \oplus \{4,5,7\} \oplus
\{3\} \oplus \{4,5,6\} \oplus \{2,3,5,6\}$\\
& & $\oplus \{2,3,4,6\} \oplus \{3,4,5,7\} = \{2,6\}$ \\
$\{J_2,J_4\}$ & $\{3,4\}$ & $\{3\} \oplus \{4,5,6\} \oplus \{2,3,5,6\} = \{2,4\}$ \\
$\{J_4,J_5\}$ & $\{4,5\}$ & $\{2,3,5,6\} \oplus \{2,3,4,6\} = \{4,5\}$\\
$\{J_5,J_7\}$ & $\{5,6,7\}$ & $\{2,3,4,6\} \oplus \{3,4,5,7\} \oplus \{2,6\} = \{5,7\}$\\
$\{J'_2,J'_7\}$ & $\{2,\ldots,7\}$ & $\{2\} \oplus \{4,5,7\} \oplus
\ldots \oplus \{2,6\} = \emptyset$
\end{tabular}

Again, we have now completely determined the reduction graph; it is
shown in Figure~\ref{ex2_fig2_red_graph}.
\end{Example}

Figures~\ref{ex_compress_realistic_graph_fig1} and
\ref{ex2_fig2_red_graph} show, for $u = 72673456 \bar 3 \bar 2 45$,
that $\compress(\fredgr_u) \approx \fredgr_\overlapgr$. The next
theorem shows that this is true for every realistic string $u$.

\begin{Theorem} \label{overlap_to_redgraph}
Let $u$ be a realistic string. Then, $\compress(\fredgr_u) \approx
\fredgr_{\overlapgru{u}}$.
\end{Theorem}
\begin{Proof}
Let $\kappa = |\dom(u)|+1$, let $\overlapgr = \overlapgru{u}$, let
$\fredgr_{\overlapgr} = (V_{\overlapgr}, E_{\overlapgr},
f_{\overlapgr}, \dom(u))$, let $R_u = \compress(\fredgr_u) = (V_u,
E_u, f_u, \dom(u))$ and let $L$ be a root subgraph of $\fredgr_u$.
Recall that the elements of $V_u$ are the desire edges of
$\fredgr_u$.

Let $h: V_u \rightarrow V_{\overlapgr}$ defined by
$$
h(v) = \begin{cases}
J_{f_u(v)} & \mbox{if $v$ is not an edge of $L$} \\
J'_{f_u(v)} & \mbox{if $v$ is an edge of $L$}
\end{cases}.
$$
We will show that $h$ is an isomorphism from $R_u$ to
$\fredgr_{\overlapgr}$. Since for every $l \in \dom(u)$ there exists
exactly one desire edge $v$ of $\fredgr_u$ that belongs to $L$ with
$f_u(v) = l$ and there exists exactly one desire edge $v$ of
$\fredgr_u$ that does not belong to $L$ with $f_u(v) = l$, it
follows that $h$ is one-to-one and onto. Also, it is clear from the
definition of $f_{\overlapgr}$ that $f_u(v) = f_{\overlapgr}(h(v))$.
Thus, it suffices to prove that $\{v_1,v_2\} \in E_u \Leftrightarrow
\{h(v_1),h(v_2)\} \in E_{\overlapgr}$.

We first prove the forward implication $\{v_1,v_2\} \in E_u
\Rightarrow \{h(v_1),h(v_2)\} \in E_{\overlapgr}$. Let $\{v_1,v_2\}
\in E_u$, let $p = f_u(v_1)$ and let $q = f_u(v_2)$. Clearly, $v_1
\not= v_2$. By the definition of $\compress$, there is a reality
edge $\tilde e = \{ \tilde v_1, \tilde v_2 \}$ of $\fredgr_u$ with
$\tilde v_1 \in v_1$ and $\tilde v_2 \in v_2$ (and thus $\tilde v_1$
and $\tilde v_2$ are labelled by $p$ and $q$ in $\fredgr_u$,
respectively). Let $i$ be the position of $\tilde e$. We consider
four cases (remember that $v_1$ and $v_2$ are both desire edges of
$\fredgr_u$):
\begin{enumerate}
\item
Assume that $\tilde e$ belongs to $L$. Then clearly, $v_1$ and $v_2$
are edges of $L$. Without loss of generality, we can assume that $p
\leq q$. From the structure of root subgraph and the fact that
$\tilde e$ is a reality edge of $\fredgr_u$ in $L$, it follows that
$q = p+1$. Now, $h(v_1) = J'_p$ and $h(v_2) = J'_q = J'_{p+1}$. By
the first item of the definition of reduction graph of an overlap
graph, it follows that $\{h(v_1), h(v_2) \} = \{J'_p, J'_{p+1}\} \in
E_{\overlapgr}$. This proves the first case. In the remaining cases,
$\tilde e$ does not belong to $L$.
\item
Assume that $v_1$ and $v_2$ are both not edges of $L$ (thus $\tilde
e$ does not belong to $L$).
%
%
Now by Theorem~\ref{th_partone_char} and the second item of the
definition of reduction graph of an overlap graph, it follows that
$\{h(v_1), h(v_2) \} = \{J_p, J_q\} \in E_{\overlapgr}$. This proves
the second case.
\item
Assume that either $v_1$ or $v_2$ is an edge of $L$ and that the
other one is not an edge of $L$ (thus $\tilde e$ does not belong to
$L$). We follow the same line of reasoning as we did in
Theorem~\ref{th_partone_char}. Without loss of generality, we can
assume that $v_1$ is not an edge of $L$ and that $v_2$ is an edge of
$L$. Clearly,
$$
\emptyset = O_u(i,i) = O_u(i,i_1) \oplus O_u(i_1,i)
$$
for each position $i_1$. By the structure of $L$ we know that $q =
2$ or $q = \kappa$. We prove it for the case $q = 2$ ($q = \kappa$,
resp.). By Lemma~\ref{overlap_edge_2} and
Lemma~\ref{overlap_edge_realistic}, we can choose $i_1 \in \{
\rspos_{p-1}, \rspos_p \}$ such that $O_u(i_1,i) = \{p\}$. By
applying Corollary~\ref{overlap_edge_1_iterative} on $L$, we have
$O_u(i,i_1) = \left( \posve(u) \cap P \right) \oplus
\left(\bigoplus_{t \in P} O_u(t)\right)$ with $P = \{2,\ldots,p-1\}
\cup P'$ ($P = \{p+1,\ldots,\kappa\} \cup P'$, resp.) for some $P'
\subseteq \{p\}$. By the third (fourth, resp.) item of the
definition of reduction graph of an overlap graph, it follows that
$\{h(v_1), h(v_2) \} = \{J'_2, J_q\} \in E_{\overlapgr}$ ($\{h(v_1),
h(v_2) \} = \{J'_{\kappa}, J_q\} \in E_{\overlapgr}$, resp.). This
proves the third case.
\item
Assume that both $v_1$ and $v_2$ are edges of $L$, but $\tilde e$
does not belong to $L$. Again, we follow the same line of reasoning
as we did in Theorem~\ref{th_partone_char}. Without loss of
generality, we can assume that $p \leq q$. By the structure of $L$,
we know that $p = 2$ and $q = \kappa > 3$. By applying
Corollary~\ref{overlap_edge_1_iterative} on $L$, we have $\emptyset
= O_u(i,i) = \left( \posve(u) \cap P \right) \oplus
\left(\bigoplus_{t \in P} O_u(t)\right)$ with $P =
\{2,\ldots,\kappa\}$. By the fifth item of the definition of
reduction graph of an overlap graph, it follows that $\{h(v_1),
h(v_2) \} = \{J'_2, J'_{\kappa}\} \in E_{\overlapgr}$. This proves
the last case.
\end{enumerate}
This proves the forward implication. We now prove the reverse
implication $\{v_1,v_2\} \in E_{\overlapgr} \Rightarrow
\{h^{-1}(v_1),h^{-1}(v_2)\} \in E_u$, where $h^{-1}$, the inverse of
$h$, is given by:
$$
\begin{array}{l}
\mbox{$h^{-1}(J_p)$ is the unique $v \in V_u$ with $f_u(v) = p$ that is not an edge of $L$,}\\
\mbox{$h^{-1}(J'_p)$ is the unique $v \in V_u$ with $f_u(v) = p$ that is an edge of $L$,}\\
\end{array}
$$
for $2 \leq p \leq \kappa$. Let $e \in E_{\overlapgr}$. We consider
each of the five types of edges in the definition of reduction graph
of an overlap graph.
\begin{enumerate}
\item
Assume $e$ is of the first type. Then $e = \{J'_p, J'_{p+1}\}$ for
some $p$ with $2 \leq p < \kappa$. Since $h^{-1}(J'_p)$ is the
desire edge of $L$ with both vertices labelled by $p$ and
$h^{-1}(J'_{p+1})$ is the desire edge of $L$ with both vertices
labelled by $p+1$, it follows, by the definition of root subgraph,
that $h^{-1}(J'_p)$ and $h^{-1}(J'_{p+1})$ are connected by a
reality edge in $L$. Thus, we have
$\{h^{-1}(J'_p),h^{-1}(J'_{p+1})\} \in E_u$. This proves the reverse
implication when $e$ is of the first type (in
Definition~\ref{def_reduction_graph_overlap_graph}).
\item
Assume $e$ is of the second type. Then $e = \{J_p, J_q\}$ for some
$p$ and $q$ with $2 \leq p < q \leq \kappa$ and
$$
\emptyset = \left( \posve(u) \cap P \right) \oplus \{p\} \oplus
\{q\} \oplus \left(\bigoplus_{t \in P} O_u(t)\right)
$$
with $P = \{p+1,\ldots,q-1\} \cup P'$ for some $P' \subseteq
\{p,q\}$.
%
%
By Theorem~\ref{th_partone_char}, there is a reality edge
$\{w_1,w_2\}$ in $\fredgr_u$, such that $w_1$ has label $p$ and
$w_2$ has label $q$ and both are not vertices of $L$. By the
definition of $\compress$, we have a $\{w'_1, w'_2\} \in E_u$ such
that $f_u(w'_1) = p$ ($f_u(w'_2) = q$, resp.) and $w'_1$ ($w'_2$,
resp.) is not an edge of $L$. Therefore $w'_1 = h^{-1}(J_p)$ and
$w'_2 = h^{-1}(J_q)$. This proves the reverse implication when $e$
is of the second type.
\item
The last three cases are proved similarly.
\end{enumerate}
This proves the reverse implication and we have shown that $h$ is an
isomorphism from $R_u$ to $\fredgr_{\overlapgr}$.
\end{Proof}
\begin{figure}
$$
\xymatrix @=30pt{
& 4 \redge[r] \dedge[d] & 7 \dedge[d] & & & 2 \dedge[d] \redge[r] & 6 \dedge[d] \\
& 4 \redge[r] & 7 & & & 2 \redge[r] & 6 \\
2 \dedge[r] & 2 \redge[r] & 3 \dedge[r] & 3 \redge[r] & 4 \dedge[r] & 4 \redge[r] & 5 \dedge[r] & 5 \redge[d] \\
3 \redge[u] \dedge[r] & 3 \redge[r] & 5 \dedge[r] & 5 \redge[r] & 7 \dedge[r] & 7 \redge[r] & 6 \dedge[r] & 6 \\
}
$$
\caption{The reduction graph of $u$ of
Example~\ref{ex3_overlap_graph}. The vertices in the figure are
represented by their labels.} \label{ex3_fig1_red_graph}
\end{figure}
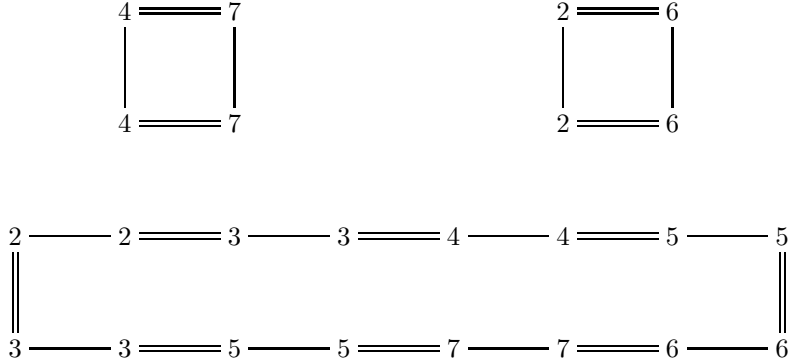
\begin{Example} \label{ex3_overlap_graph}
The realistic string $u = 453475623267$ was introduced in
Example~\ref{ex1_overlap_graph}. The reduction graph
$\fredgr_{\overlapgr}$ of the overlap graph of $u$ is given in
Figure~\ref{ex1_fig2_red_graph}. The reduction graph $\fredgr_u$ of
$u$ is given in Figure~\ref{ex3_fig1_red_graph}. It is easy to see
that the result of applying $\compress$ to $\fredgr_u$ is a graph
that is indeed isomorphic to $\fredgr_{\overlapgr}$. This makes
clear why there were two proofs for both edges $\{J_2,J_6\}$ and
$\{J_4,J_7\}$ in Example~\ref{ex1_overlap_graph}; each one
corresponds to one reality edge in $\fredgr_u$ (outside $L$).
\end{Example}
%
%
Formally, we have not yet (up to isomorphism) constructed the
reduction graph $\fredgr_u$ of a realistic string $u$ from its
overlap graph. We have `only' constructed $\compress(\fredgr_u)$ (up
to isomorphism). However, it is clear that $\fredgr_u$ can easily be
obtained from $\compress(\fredgr_u)$ (up to isomorphism) by
considering the edges as reality edges and replacing every vertex by
a desire edge of the same label.

\section{Consequences} \label{sect_consequences}
Using the previous theorem and \cite{SuccessfulnessChar_Original}
(or Chapter~13 in \cite{GeneAssemblyBook}), we can now easily
characterize successfulness for realistic overlap graphs in any
given $S \subseteq \{Gnr,Gpr,Gdr\}$. The notions of successful
reduction, string negative rule and graph negative rule used in this
section are defined in \cite{GeneAssemblyBook}.

Below we restate a theorem of \cite{Extended_paper}.
\begin{Theorem} \label{th_recall_number_cc}
Let $N$ be the number of components in $\fredgr_u$. Then every
successful reduction of $u$ has exactly $N-1$ string negative rules.
\end{Theorem}
Due to the `weak equivalence' of the string pointer reduction system
and the graph pointer reduction system, proved in Chapter 11 of
\cite{GeneAssemblyBook}, we can, using
Theorem~\ref{overlap_to_redgraph}, restate
Theorem~\ref{th_recall_number_cc} in terms of graph reduction rules.
\begin{Theorem} \label{char_gnr_overlap_graph}
Let $u$ be a realistic string, and $N$ be the number of components
in $\fredgr_{\overlapgru{u}}$. Then every successful reduction of
$\overlapgru{u}$ has exactly $N-1$ graph negative rules.
\end{Theorem}
As an immediate consequence we have the following corollary. It
provides a solution to an open problem formulated in Chapter~13 in
\cite{GeneAssemblyBook}.
\begin{Corollary} \label{char_successfulness_overlap_graph}
Let $u$ be a realistic string. Then $\overlapgru{u}$ is successful
in $\{Gpr,Gdr\}$ iff $\fredgr_{\overlapgru{u}}$ is connected.
\end{Corollary}

\begin{Example}
Every successful reduction of the overlap graph of
Example~\ref{ex1_overlap_graph} has exactly two graph negative
rules, because its reduction graph consist of exactly three
components. For example ${\bf gnr}_4 \ {\bf gdr}_{5,7} \ {\bf gnr}_2
\ {\bf gdr}_{3,6}$ is a successful reduction of this overlap graph.

Every successful reduction of the overlap graph of
Example~\ref{ex2_overlap_graph} has exactly one graph negative rule.
For example ${\bf gnr}_2 \ {\bf gpr}_4 \ {\bf gpr}_5 \ {\bf gpr}_{7}
\ {\bf gpr}_6 \ {\bf gpr}_{3}$ is a successful reduction of this
overlap graph.
\end{Example}

With the help of \cite{SuccessfulnessChar_Original} (or Chapter~13
in \cite{GeneAssemblyBook}) and
Corollary~\ref{char_successfulness_overlap_graph}, we are ready to
complete the characterization of successfulness for realistic
overlap graphs in any given $S \subseteq \{Gnr, Gpr, Gdr\}$.

\begin{Theorem} \label{th_char}
Let $u$ be a realistic string. Then $\overlapgru{u}$ is successful
in:
\begin{itemize}
\item
$\{Gnr\}$ iff $\overlapgru{u}$ is a discrete graph with only
negative vertices.
\item
$\{Gnr,Gpr\}$ iff each component of $\overlapgru{u}$ that consists
of more than one vertex contains a positive vertex.
\item
$\{Gnr, Gdr\}$ iff all vertices of $\overlapgru{u}$ are negative.
\item
$\{Gnr, Gpr, Gdr\}$.
\item
$\{Gdr\}$ iff all vertices of $\overlapgru{u}$ are negative and
$\fredgr_{\overlapgru{u}}$ is connected.
\item
$\{Gpr\}$ iff each component of $\overlapgru{u}$ contains a positive
vertex and $\fredgr_{\overlapgru{u}}$ is connected.
\item
$\{Gpr,Gdr\}$ iff $\fredgr_{\overlapgru{u}}$ is connected.
\end{itemize}
\end{Theorem}

\section{Discussion} \label{sect_discussion}
We have shown how to directly construct the reduction graph of a
realistic string $u$ (up to isomorphism) from the overlap graph
$\overlapgr$ of $u$. From a biological point of view, this allows
one to reconstruct a representation of the macronuclear gene (and
its waste products) given only the overlap graph of the micronuclear
gene. Moreover, this results allows one to (directly) determine the
number $n$ of graph negative rules that are necessary to reduce
$\overlapgr$ successfully. Along with some results in previous
papers, it also allows us to give a complete characterization of the
successfulness of $\overlapgr$ in any given $S \subseteq
\{Gnr,Gpr,Gdr\}$.

It remains an open problem to find a (direct) method to determine
this number $n$ for overlap graphs $\overlapgr$ in general (not just
for realistic overlap graphs). That is, a better method than first
determining a legal string $u$ corresponding with $\overlapgr$ and
then determining the reduction graph of $u$.

\paper{
\bibliographystyle{plain}
\bibliography{../geneassembly}

\begin{thebibliography}{1}

\bibitem{DBLP:conf/complife/BrijderHR06}
R.~Brijder, H.J. Hoogeboom, and M.~Muskulus.
\newblock Applicability of loop recombination in ciliates using the breakpoint
  graph.
\newblock In M.R. Berthold et~al., editors, {\em CompLife '06}, volume 4216 of
  {\em LNCS}, pages 97--106. Springer, 2006.

\bibitem{DBLP:conf/complife/BrijderHR05}
R.~Brijder, H.J. Hoogeboom, and G.~Rozenberg.
\newblock The breakpoint graph in ciliates.
\newblock In M.R. Berthold et~al., editors, {\em CompLife '05}, volume 3695 of
  {\em LNCS}, pages 128--139. Springer, 2005.

\bibitem{Extended_paper}
R.~Brijder, H.J. Hoogeboom, and G.~Rozenberg.
\newblock Reducibility of gene patterns in ciliates using the breakpoint graph.
\newblock {\em Theor. Comput. Sci.}, 356:26--45, 2006.

\bibitem{GeneAssemblyBook}
A.~Ehrenfeucht, T.~Harju, I.~Petre, D.M. Prescott, and G.~Rozenberg.
\newblock {\em Computation in Living Cells -- Gene Assembly in Ciliates}.
\newblock Springer Verlag, 2004.

\bibitem{SuccessfulnessChar_Original}
A.~Ehrenfeucht, T.~Harju, I.~Petre, and G.~Rozenberg.
\newblock Characterizing the micronuclear gene patterns in ciliates.
\newblock {\em Theory of Computing Systems}, 35:501--519, 2002.

\bibitem{DBLP:conf/birthday/HarjuPR04}
T.~Harju, I.~Petre, and G.~Rozenberg.
\newblock Formal properties of gene assembly: Equivalence problem for overlap
  graphs.
\newblock In N.~Jonoska, G.~Paun, and G.~Rozenberg, editors, {\em Aspects of
  Molecular Computing}, volume 2950 of {\em LNCS}, pages 202--212. Springer,
  2004.

\end{thebibliography}

\end{document}
}